\documentclass[manuscript]{aastex}
\usepackage{xcolor}

\shorttitle{Temporal variations of multi-TeV cosmic ray anisotropy}
\shortauthors{The Tibet AS$\gamma$ Collaboration}

\begin{document}

\title{On temporal variations of the multi-TeV cosmic ray anisotropy
   using the Tibet III Air Shower Array}

\author{
    M.~Amenomori\altaffilmark{1}, X.~J.~Bi\altaffilmark{2},
    D.~Chen\altaffilmark{3}, S.~W.~Cui\altaffilmark{4},
    Danzengluobu\altaffilmark{5}, L.~K.~Ding\altaffilmark{2},
        X.~H.~Ding\altaffilmark{5}, C.~Fan\altaffilmark{6}\,\altaffilmark{2},
    C.~F.~Feng\altaffilmark{6}, Zhaoyang~Feng\altaffilmark{2},
    Z.~Y.~Feng\altaffilmark{7}, X.~Y.~Gao\altaffilmark{8},
    Q.~X.~Geng\altaffilmark{8},
        Q.~B.~Gou\altaffilmark{2}, H.~W.~Guo\altaffilmark{5},
    H.~H.~He\altaffilmark{2}, M.~He\altaffilmark{6},
    K.~Hibino\altaffilmark{9}, N.~Hotta\altaffilmark{10},
    Haibing~Hu\altaffilmark{5}, H.~B.~Hu\altaffilmark{2},
        J.~Huang\altaffilmark{2}, Q.~Huang\altaffilmark{7},
    H.~Y.~Jia\altaffilmark{7}, L.~Jiang\altaffilmark{8}\,\altaffilmark{2},
    F.~Kajino\altaffilmark{11}, K.~Kasahara\altaffilmark{12},
    Y.~Katayose\altaffilmark{13},
        C.~Kato\altaffilmark{14}, K.~Kawata\altaffilmark{3},
    Labaciren\altaffilmark{5}, G.~M.~Le\altaffilmark{15},
    A.~F.~Li\altaffilmark{6}, H.~C.~Li\altaffilmark{4}\,\altaffilmark{2},
    J.~Y.~Li\altaffilmark{6}, C.~Liu\altaffilmark{2},
        Y.-Q.~Lou\altaffilmark{16}, H.~Lu\altaffilmark{2},
    X.~R.~Meng\altaffilmark{5}, K.~Mizutani\altaffilmark{12}\,\altaffilmark{17},
    J.~Mu\altaffilmark{8}, K.~Munakata\altaffilmark{14},
    A.~Nagai\altaffilmark{18},
        H.~Nanjo\altaffilmark{1}, M.~Nishizawa\altaffilmark{19},
    M.~Ohnishi\altaffilmark{3}, I.~Ohta\altaffilmark{20},
    S.~Ozawa\altaffilmark{12}, T.~Saito\altaffilmark{21},
    T.~Y.~Saito\altaffilmark{22},
        M.~Sakata\altaffilmark{11}, T.~K.~Sako\altaffilmark{3},
    M.~Shibata\altaffilmark{13}, A.~Shiomi\altaffilmark{23},
    T.~Shirai\altaffilmark{9}, H.~Sugimoto\altaffilmark{24},
    M.~Takita\altaffilmark{3},
        Y.~H.~Tan\altaffilmark{2}, N.~Tateyama\altaffilmark{9},
    S.~Torii\altaffilmark{12}, H.~Tsuchiya\altaffilmark{25},
    S.~Udo\altaffilmark{9}, B.~Wang\altaffilmark{2},
    H.~Wang\altaffilmark{2}, Y.~Wang\altaffilmark{2},
        Y.~G.~Wang\altaffilmark{6}, H.~R.~Wu\altaffilmark{2},
    L.~Xue\altaffilmark{6}, Y.~Yamamoto\altaffilmark{11},
    C.~T.~Yan\altaffilmark{26}, X.~C.~Yang\altaffilmark{8},
    S.~Yasue\altaffilmark{27},
        Z.~H.~Ye\altaffilmark{28}, G.~C.~Yu\altaffilmark{7},
    A.~F.~Yuan\altaffilmark{5}, T.~Yuda\altaffilmark{9},
    H.~M.~Zhang\altaffilmark{2}, J.~L.~Zhang\altaffilmark{2},
    N.~J.~Zhang\altaffilmark{6},
        X.~Y.~Zhang\altaffilmark{6}, Y.~Zhang\altaffilmark{2},
    Yi~Zhang\altaffilmark{2}, Ying~Zhang\altaffilmark{7}\,\altaffilmark{2},
    Zhaxisangzhu\altaffilmark{5} and X.~X.~Zhou\altaffilmark{7}\\
        (The Tibet AS$\gamma$ Collaboration)
}


\altaffiltext{1}{Department of Physics, Hirosaki University, Hirosaki 036-8561, Japan.}
\altaffiltext{2}{Key Laboratory of Particle Astrophysics, Institute of High Energy Physics, Chinese Academy of Sciences, Beijing 100049, China.}
\altaffiltext{3}{Institute for Cosmic Ray Research, University of Tokyo, Kashiwa 277-8582, Japan.}
\altaffiltext{4}{Department of Physics, Hebei Normal University, Shijiazhuang 050016, China.}
\altaffiltext{5}{Department of Mathematics and Physics, Tibet University, Lhasa 850000, China.}
\altaffiltext{6}{Department of Physics, Shandong University, Jinan 250100, China.}
\altaffiltext{7}{Institute of Modern Physics, SouthWest Jiaotong University, Chengdu 610031, China.}
\altaffiltext{8}{Department of Physics, Yunnan University, Kunming 650091, China.}
\altaffiltext{9}{Faculty of Engineering, Kanagawa University, Yokohama 221-8686, Japan.}
\altaffiltext{10}{Faculty of Education, Utsunomiya University, Utsunomiya 321-8505, Japan.}
\altaffiltext{11}{Department of Physics, Konan University, Kobe 658-8501, Japan.}
\altaffiltext{12}{Research Institute for Science and Engineering, Waseda University, Tokyo 169-8555, Japan.}
\altaffiltext{13}{Faculty of Engineering, Yokohama National University, Yokohama 240-8501, Japan.}
\altaffiltext{14}{Department of Physics, Shinshu University, Matsumoto 390-8621, Japan.}
\altaffiltext{15}{National Center for Space Weather, China Meteorological Administration, Beijing 100081, China.}
\altaffiltext{16}{Physics Department and Tsinghua Center for Astrophysics, Tsinghua University, Beijing 100084, China.}
\altaffiltext{17}{Saitama University, Saitama 338-8570, Japan.}
\altaffiltext{18}{Advanced Media Network Center, Utsunomiya University, Utsunomiya 321-8585, Japan.}
\altaffiltext{19}{National Institute of Informatics, Tokyo 101-8430, Japan.}
\altaffiltext{20}{Sakushin Gakuin University, Utsunomiya 321-3295, Japan.}
\altaffiltext{21}{Tokyo Metropolitan College of Industrial Technology, Tokyo 116-8523, Japan.}
\altaffiltext{22}{Max-Planck-Institut f\"ur Physik, M\"unchen D-80805, Deutschland.}
\altaffiltext{23}{College of Industrial Technology, Nihon University, Narashino 275-8576, Japan.}
\altaffiltext{24}{Shonan Institute of Technology, Fujisawa 251-8511, Japan.}
\altaffiltext{25}{RIKEN, Wako 351-0198, Japan.}
\altaffiltext{26}{Institute of Disaster Prevention Science and Technology, Yanjiao 065201, China.}
\altaffiltext{27}{School of General Education, Shinshu University, Matsumoto 390-8621, Japan.}
\altaffiltext{28}{Center of Space Science and Application Research, Chinese Academy of Sciences, Beijing 100080, China.}


\begin{abstract}
We analyze the large-scale two-dimensional sidereal anisotropy of
multi-TeV cosmic rays by Tibet Air Shower Array, with the data
taken from 1999 November to 2008 December. To explore temporal
variations of the anisotropy, the data set is divided into nine
intervals, each in a time span of about one year. The sidereal
anisotropy of magnitude about 0.1\% appears fairly stable from
year to year over the entire observation period of nine years.
This indicates that the anisotropy of TeV Galactic cosmic rays
remains insensitive to solar activities since the observation
period covers more than a half of the 23rd solar  cycle.
\end{abstract}

\section{Introduction}

Galactic cosmic rays (GCRs) are high-energy nuclei (most protons)
which are believed to be accelerated by supernova remnants (SNRs) in our
Galaxy and continuously reach the Earth after propagating in the Galaxy
and heliosphere. The intensity of GCRs is nearly isotropic due to
deflections in the Galactic magnetic field (GMF). However,
extensive observations do show that there exists a slight
anisotropy on the overall isotropic background
\citep[e.g.][]{Jacklyn1966,Nagashima1975,Cutler1981,Nagashima1989,
Aglietta1996,Munakata1997,sci2006,Milagro2009,IceCube2009}.

The large-scale sidereal cosmic ray (CR) anisotropy may arise from
several causes. Firstly, it may result from the uneven distribution
of CR sources such as SNRs and the process of
CR propagation in the Galaxy. 
Using the diffusion model, it is expected that the amplitude of
anisotropy satisfies $A \sim D(R) \sim R^{1/3}$, where $D(R)$ is
the diffusion coefficient which depends on the rigidity $R$ of
charged particles \citep[e.g.][]{Berezinskii90ICRC,Shibata2004}.
Thus one can expect that the anisotropy amplitude will increase with
energy. However, the observation of the anisotropy above hundreds of TeV
is smaller than what is expected.
It has been suggested that a nearby (within 1-2 kpc from the Sun)
source in the opposite direction against the global CR diffusion may
suppress the expected large amplitude to some degree \citep{Erlykin2006}.
Other factors such as the complicated composition of CRs and
isotropic CRs from the Galactic halo can also be helpful for reducing the
anisotropy \citep[e.g.][]{Erlykin2006,Ptuskin2006}.
Secondly, the anisotropy can also be induced through both large-scale and local
magnetic field configurations, possibly including effects of the
heliosphere. Therefore, it is a useful tool to probe the local
interstellar space surrounding the heliosphere and the magnetic
structure of the heliosphere
\citep{ImplicationAIP2007,Munakata-arxiv2008,Amenomori09ICRCaniso}.
In addition, an expected anisotropy is caused by the relative
motion between the observer and the CR plasma, known as
the Compton-Getting (CG) effect \citep{CG1935}.
By analyzing the events recorded by the Tibet III Air Shower
experiment, \citet{sci2006} showed that GCRs corotate with the
local GMF environment according to the null result of the Galactic CG
effect. The CG modulation due to the Earth's orbital motion around
the Sun has been successfully detected by several GCR experiments
in multi-TeV energy ranges
\citep[e.g.][]{Nat1986,Amenomori04PRL,sci2006,Milagro2009}.

From the analysis of numerous experiments, it can be seen that
both the amplitude and the phase of the best-fit first harmonic
vary with CR energy in a wide range from tens of GeV to PeV
\citep{Guillian2007,IceCube2009}. Below several tens of GeV, solar
modulation effects are most notable for GCRs. GCRs interact with
the solar wind magnetic field, both an ordered field and
irregular field components, after entering the heliosphere. The spatial
distribution of GCRs can reflect the magnetic structure in the solar
wind. With increasing energy, CRs become less sensitive to
the solar modulation. It is well known that the flux of GCRs
with energy per nucleon in the energy range of $\sim 10^{11}
-10^{14}$ eV has a sidereal anisotropy of the order $O(10^{-3})$.
The gyro radius $r_L$ of CRs in this energy range in a GMF
of 3 $\mu$G is about several AU$-$0.03 pc, which is much smaller
than the size of the Galaxy. In the multi-TeV range, the gyro radius of
hundreds of AU becomes comparable to the spatial scale of the heliosphere in
the nose direction toward the upstream side of the interstellar
medium flow \citep[e.g.][]{Washimi1996}. However, it is known that
the heliosphere has a long heliotail, the modulation in the
heliotail remains possible. Therefore, the large-scale
sidereal anisotropy of CRs in this energy range gives us an
important clue about the magnetic field structure of the
heliosphere or the local interstellar space surrounding the
heliosphere.

The solar cycle shows
a quasi-period of about 11 yr and the global magnetic polarity
reverses with a quasi-period of two solar activity cycles
\citep[e.g.][]{Solanki2006}. Since GCRs are  modulated by solar
activities in the given energy range mentioned above, it might be
expected that the sidereal anisotropy may follow the variation of
solar cycle. There have been many experiments devoted to the study
of temporal variations of GCR sidereal anisotropy.
\citet{Nagashima1989} found that both the amplitude and the phase of the
sidereal anisotropy with the rigidity $\sim 10^{13}$V had no significant
changes for more than ten years, except that the phase changed at the epoch
 (1979-1980) when the polar magnetic field of the Sun reversed its polarity.
At primary energy as low as 2 TeV, \citet{Baksan1995} demonstrated that the 
sidereal anisotropy was constant within the accuracy of the experiment, with
observations in the subsequent ten years (1982-1991) after this epoch of polarity reversal.
Recently, underground muon observations showed that yearly mean harmonic vectors
of the sidereal anisotropy in the sub-TeV region are more or less stable in phase
and amplitude over 20 years since 1985 without showing any significant correlation
with the solar activity and magnetic-cycles \citep{Munakata-arxiv2008}.
However, Milagro experiment recently  reported an amplitude increase of
the sidereal anisotropy at 6 TeV in the latter half of the 23rd solar
cycle (from 2000 July to 2007 July), while the phase remains stable
\citep{Milagro2009} . The contradiction needs further checks from other
experiments.

Covering almost the same field of view and with similar sensitive
energy range, the Tibet AS$\gamma$ experiment seemingly observed
no variation in the sidereal anisotropy using data from both Tibet
HD and Tibet III array which was divided into two parts by the
solar maximum around 2001 \citep{sci2006}. In the multi-TeV
region, \citet{Amenomori05APJ} found that both the amplitude and
phase of the first harmonic vector of the daily variation are
remarkably independent of primary energy.
Taking advantage of the large field of view and high count rates as
well as the good angular resolution of the incident direction,
the Tibet III Air Shower Array provides currently the world's
highest precision two-dimensional measurement of GCR intensity in
this energy range. The observation period runs from 1999 November
to 2008 December, covering more than a half of the 23rd solar
activity cycle from the maximum to the minimum. Therefore, we can
do more precise study on the sidereal anisotropy variation year by
year in correlation with the solar activity cycle.
In this paper, we analyze temporal variations of sidereal
anisotropy of multi-TeV GCR intensity using the data of Tibet III
array from 1999 November to 2008 December.
At this point,
we cannot investigate the influence of the polarity reversal of
the global solar magnetic field on the sidereal anisotropy due to
the lack of data before the current magnetic field reversal.
Note that we do not use the data of Tibet HD to perform the
year-by-year analysis due to the limitation of the statistics,
although it covers the period before the reversal.

\section{Tibet Air Shower Array Experiment}
\label{experiment}

The Tibet Air Shower Array experiment has been
operating successfully at Yangbajing (90.522$^\circ$ E,
30.102$^\circ$ N; 4300 m above the sea level) in Tibet, China
since 1990. The array has been gradually upgraded by increasing
the number of counters from the Tibet-I array with 65 plastic
scintillation detectors placed on a lattice with a 15 m spacing
\citep{TB1992}. The Tibet III array was completed in the late fall
of 1999 \citep{TB2001}.  The array is composed of 497 fast timing
(FT) detectors and 36 density (D) detectors, covering a surface
area of 22,050 m$^2$. Each FT detector equipped with a plastic
scintillator plate and a 2 inch photomultiplier tube has a cross
sectional area of 0.5 m$^2$ and is deployed at a lattice with a
7.5 m spacing. A 0.5 cm thick lead plate is placed on top of each
counter in order to increase the detector array sensitivity by
converting $\gamma$ rays into electron-positron pairs. A CR event
trigger signal is issued when any fourfold coincidence occurs in
the FT counters recording more than 0.6 particles, resulting in a
trigger rate of about 680 Hz at a few-TeV threshold energy. The
shower size $\sum \rho_{\rm {FT}}$ is regarded as an estimator for
the primary particle energy, where the size of $\sum \rho_{\rm
{FT}}$ is defined as the sum of particles per m$^2$ for each FT
detector. During 2002 and 2003, the inside area of the Tibet-III array
was further enlarged to 36,900 m$^2$ by installing additional 256
detectors. This full Tibet III array has been operating
successfully since 2003. In the present analysis, to keep the
uniformity of the data quality throughout the entire observation
period from 1999 to 2008, we reconstructed CR air shower events
obtained using the configuration of detectors which was completed
in the late fall 1999 even for the full Tibet-III array.

In the present analysis, CR events are selected based on the
following four criteria: (1) estimated air shower core location
should be inside the array; (2) the zenith angle of the incident
direction should be less than $45^\circ$; (3) any fourfold
coincidence in the FT counters should record a signal of more than
0.8 particles; (4) when $10 \leq \sum \rho_{\rm {FT}} < 178$ is
satisfied, corresponding to a modal energy of about 5 TeV.

In total, about $4.91\times10^{10}$ CR events are used in the
present analysis.

\section{Analysis and Results}
Sitting on an almost horizontal plane, the Tibet III Air Shower
Array has almost azimuth-independent efficiency in receiving GCR
shower events for any given zenith angle. Therefore, the
equi-zenith angle method is adopted in the present analysis. In
brief, for a candidate ``on-source window'', GCR air shower events
recorded simultaneously in the side band of the same
zenith angle belt can be used to construct the ``off-source
windows'' and to estimate the background events.
This method can eliminate various detection effects caused by
instrumental and environmental variations, such as changes in
pressure and temperature which are hard to be controlled and
tend to introduce systematic errors in measurement.

With the large statistics, we can construct a two-dimensional sky
map to reveal detailed structural information of the large-scale
GCR variations beyond the simple one-dimensional profile along the
time scale. The idea of this method is that for each short time
step (e.g. 8 minutes in the present analysis), for all directions,
if we scale down (or up) the number of observed events by dividing
them according to their relative CR intensity, then statistically,
those scaled observed numbers of events in a zenith angle belt
should be equal anywhere. A total $\chi^2$ can be built accordingly,
and the relative intensity of CRs $I$ and its error
$\Delta I$ in each direction can be solved by minimizing the
$\chi^2$ function (see \citet{All-Sky} for more details).
Given the fact that the detector can quickly scan the sky along
right ascension direction as a result of the self rotation of the 
Earth, the intensities of CRs along right ascension direction are 
measured under the same condition for each declination belt. Therefore, 
the modulation along right ascension direction in each declination belt 
is accurate and irrelevant to the inhomogeneous efficiency of the detector.
However, as having been pointed out by \citep[e.g.][]{Andreyev2008,Kozyarivsky2008},
when comparing CR intensities among different declination belts, we need absolute 
efficiency calibration along declination direction which is unfortunately not possible 
for current experiments. Once we have the absolute CR intensity in declination direction 
$I_{dec,1D}(\delta)$, we will obtain the true two-dimensional anisotropy map, 
$I_{true,2D}(\alpha,\delta)=I_{quasi,2D}(\alpha,\delta)\times I_{dec,1D}(\delta)$,
here $\alpha$ and $\delta$ are right ascension and declination in the celestial coordinate 
respectively. Strictly speaking, the present analysis gives a combination of one-dimensional 
modulation curves in a fine set of declination bins -- $I_{quasi,2D}(\alpha,\delta)$ (For 
simplicity, we will omit the subscript of $I_{quasi,2D}$ in the following text).

Using Lomb-Scargle Fourier transformation method \citep{Lomb1976,Scargle1982}
with CR data recorded by the Tibet III array, Tibet AS$\gamma$ experiment 
showed that besides the well-known solar diurnal, sidereal diurnal and sidereal 
semi-diurnal modulations at a level of $\sim 10^{-3}$, no other periodicity was 
found to have high enough significance from 1 hour to 2 years in the energy range 
from $\sim 3.0$ TeV to $\sim 12.0$ TeV \citep{PeriodicLiaf}. The sidereal daily 
variation can be described by the first and second harmonics and the solar daily 
variation can be described by the first harmonic alone.
When applying the above mentioned method to data, the fit function is natural to 
contain both sidereal time and solar time modulation components. The other 
alternative method we used to separate modulations was to properly fold the data 
according to the sidereal time or solar time periodicity. This method was used 
in \citep{All-Sky,sci2006} when we analyzed CR intensity variations in the sidereal 
time frame with several years data samples. Using this method, each event had to 
be re-weighted to form an exactly one year long and uniformly distributed data 
series  before folding. Therefore, the disadvantage of this method is that more 
than one year's data are needed, which can not be followed in current work. In 
this case, we adopt a new method of fitting all modulation components simultaneously 
to perform the year-by-year analysis of CR anisotropy.

Based on the results of Lomb-Scargle Fourier transformation, in the present analysis, 
we assume that at any moment $t$, the relative intensity of CRs at any given 
direction $(\theta,\phi)$ in the horizontal coordinate, is modulated as a product 
of $I_{sid}(\alpha_{sid},\delta_{sid})$ and $I_{sol}(\alpha_{sol},\delta_{sol})$. 
Here $(\alpha_{sid},\delta_{sid})$ and $(\alpha_{sol},\delta_{sol})$ are positions 
corresponding to the celestial coordinate in the local sidereal time frame and the 
local solar time frame of the same point $(t,\theta,\phi)$ in horizontal coordinate, 
$I_{sid}$ and $I_{sol}$ denote the CR intensity in the sidereal time frame and the 
solar time frame respectively. So substituting $I$ by 
$I_{sid}(\alpha_{sid},\delta_{sid})\times I_{sol}(\alpha_{sol},\delta_{sol})$,
the total $\chi^2$ can be written as
\begin{eqnarray}
\chi^2 = & \sum_{t,\theta,\phi} \left( \left\{
\frac{N_{obs}(t,\theta,\phi)}{I_{sid}(\alpha_{sid},\delta_{sid})
\times I_{sol}(\alpha_{sol},\delta_{sol})} - \frac{\sum_{\phi\neq\phi'}
\left[N_{obs}(t,\theta,\phi')/\left(I_{sid}(\alpha_{sid}',\delta_{sid}')
\times I_{sol}(\alpha_{sol}',\delta_{sol}') \right)  \right]}
{\sum_{\phi\neq\phi'}1} \right\}^2 \right. \nonumber \\
& \times \left. \left\{ \frac{N_{obs}(t,\theta,\phi)}
{I_{sid}^2(\alpha_{sid},\delta_{sid}) \times I_{sol}^2(\alpha_{sol},\delta_{sol})} +
\frac{\sum_{\phi\neq\phi'}\left[N_{obs}(t,\theta,\phi')
/\left(I_{sid}^2(\alpha_{sid}',\delta_{sid}') \times I_{sol}^2(\alpha_{sol}',
\delta_{sol}')\right)\right]}{(\sum_{\phi\neq\phi'}1)^2}\right\}^{-1}\right)\
. \label{newchisq}
\end{eqnarray}
Here, $N_{obs}(t,\theta,\phi)$ denotes the number of observed
events in the ``window'' $(\theta, \phi)$ in the horizontal
coordinate at the moment $t$. After minimizing the $\chi^2$
function, $I_{sid}(\alpha_{sid},\delta_{sid})$ and
$I_{sol}(\alpha_{sol},\delta_{sol})$ can be obtained simultaneously.

\begin{figure}[!htb]
\centering
\includegraphics[width=0.9\columnwidth]{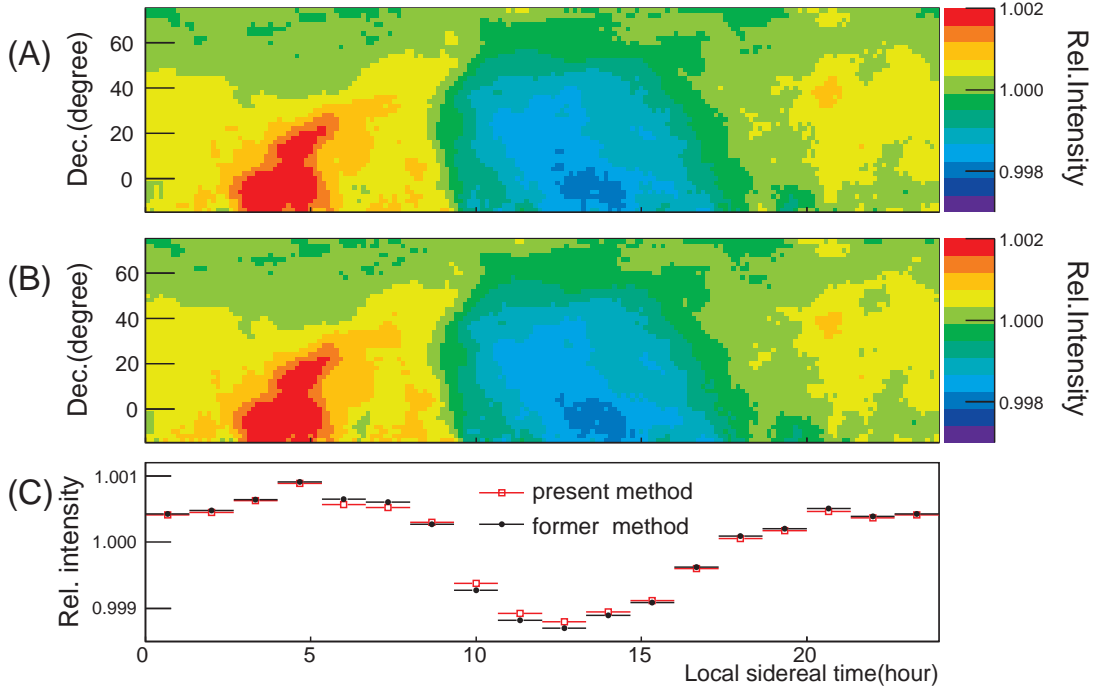}
\caption{ Sidereal diurnal variation of CR relative
   intensity with the representative primary energy of 5 TeV
   averaged over all nine phases of Tibet III Air Shower Array
   from Nov. 1999 to Dec. 2008. \ \ (A) 2D maps obtained with
   the present method mentioned as Equation (\ref{newchisq});
   (B) 2D maps obtained with the former method used in \citep{sci2006};
   (C) the 1D projection of 2D maps averaged over all declinations
   for comparison. } \label{mapsid}
\end{figure}

The data were acquired by the Tibet III Air Shower Array for
1915.5 live days from Nov. 1999 to Dec. 2008. It covers all nine
running phases of the Tibet III Air Shower Array. As described in
Equation (\ref{newchisq}), for any direction in the horizontal
coordinate at any moment in the observation period, a $\chi^2$
function can be constructed. Covering all nine phases of Tibet III
array, we got a total $\chi^2$ accumulated over the entire
observation period. By minimizing it, the CR intensity maps in the
frame of the local sidereal time and the local solar time averaged
over all nine phases were obtained simultaneously. Fig.
\ref{mapsid} (A) shows the intensity map of GCRs with the modal
energy of 5 TeV in the local sidereal time frame averaged over the
whole observation period of Tibet III array. The result is
consistent with former observation results of Tibet III experiment
by different methods
\citep{Amenomori05APJ,sci2006,ImplicationAIP2007}. The amplitude
of the sidereal anisotropy is about 0.1\%, and the maximum around
6 hr in the local sidereal time. The so-called ``tail-in'' and
``loss-cone'' anisotropy components
\citep[e.g.][]{Nagashima1976,Nagashima1998} are clearly shown in
the map. The maximum phase shifts to earlier hours as the viewing
declination moves southward in two-dimensional intensity map. The
excess around the Cygnus arm direction can also be seen in the map.

For a detailed examination between the present method and the
former method used in \citet{sci2006}, we constructed a
two-dimensional map of CR relative intensity with former method
used in \citet{sci2006} with the same data samples, showed in Fig.
\ref{mapsid} (B). The 1D projections of Fig. \ref{mapsid} (A) and
(B) averaged over all declination belts ($-14.89^\circ$ to
$75.11^\circ$) are displayed in Fig. \ref{mapsid} (C). From the
comparison, it can be seen that the results agree with each other
well.

To study the temporal variation of the sidereal anisotropy, the
data was divided into nine subsets, corresponding to nine running
phases of Tibet III array each in a time scale of about one year,
as summarized in Table \ref{phaseDiv}.

\begin{table}[!htb]
\caption{Definition of nine phases of Tibet III from 1999 November
to 2008 December} \label{phaseDiv} \centering
\begin{tabular}{ c c c c c }
\hline
   Phase  & Start time  & End time      & Live days   & Number of used CR events \\
   \hline
    1     & Nov. 18, 1999 & Jun. 29, 2000   & 173.1  &  5.16$\times 10^9$ \\
    2     & Oct. 28, 2000 & Oct. 11, 2001   & 283.7  &  8.14$\times 10^9$ \\
    3     & Dec. 05, 2001 & Sep. 19, 2002   & 201.8  &  5.59$\times 10^9$ \\
    4     & Nov. 18, 2002 & Nov. 18, 2003   & 259.1  &  6.34$\times 10^9$ \\
    5     & Dec. 14, 2003 & Oct. 10, 2004   & 123.6  &  3.07$\times 10^9$ \\
    6     & Oct. 19, 2004 & Nov. 15, 2005   & 277.6  &  6.79$\times 10^9$ \\
    7     & Dec. 07, 2005 & Nov. 03, 2006   & 114.5  &  2.71$\times 10^9$ \\
    8     & Nov. 06, 2006 & Feb. 25, 2008   & 269.2  &  6.36$\times 10^9$ \\
    9     & Mar. 02, 2008 & Dec. 03, 2008   & 212.9  &  4.91$\times 10^9$ \\
\hline
\end{tabular}
\end{table}

\begin{figure*}[!htb]
\centering
\includegraphics[width=\columnwidth]{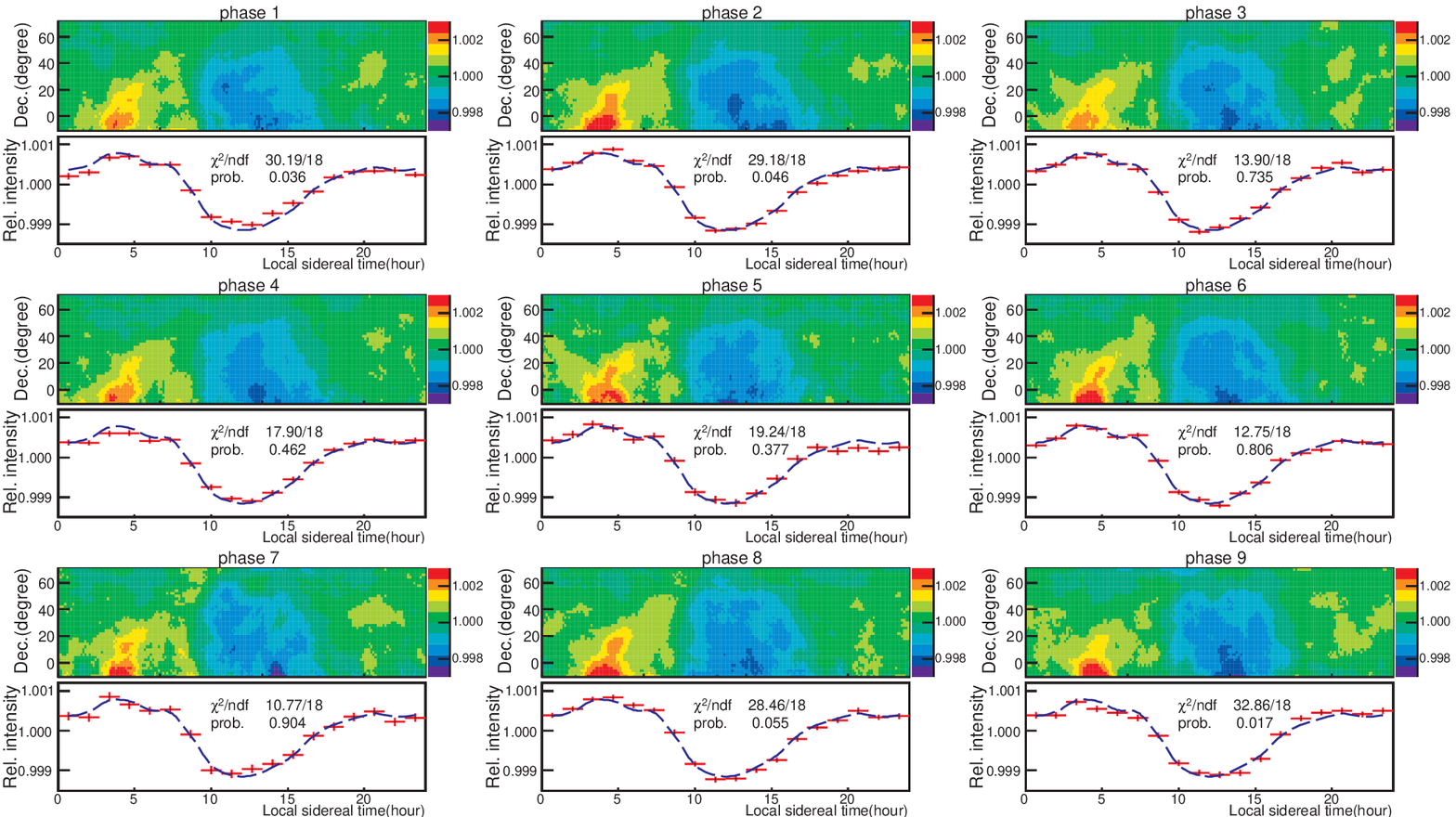}
\caption{ Cosmic ray intensity variation in the local sidereal time frame for
          CRs with the modal energy around 5 TeV in the 9 phases of Tibet III Array.
    Top: 2D intensity map of each phase; Bottom: 1D projection averaged over all 
    declinations.  In bottom plots of each panel, the red crosses in each plot show 
    the intensity variation over each phase respectively, while the dashed blue lines 
    represent the intensity averaged over all nine phases of Tibet III array.}
\label{diffphase}
\end{figure*}

Using data samples recorded during each separated phase, the GCR
relative intensity maps in the local sidereal time frame averaged
over each phase were obtained according to Equation
(\ref{newchisq}). The two-dimensional (2D) relative intensity maps
of GCRs with modal energy around 5 TeV in the local sidereal time
frame of different phases are displayed in Fig. \ref{diffphase},
and the corresponding one-dimensional (1D) projections over all
declinations are also shown. The solid red markers in each plot
denote the relative intensities of GCRs in the local sidereal time
frame over the corresponding observation period, while the blue
dashed smooth curves in plots represent variations of GCR relative
intensity averaged over all nine phases of Tibet III array, the
same as the bottom plot with the marker of ``present method'' in
Fig. \ref{mapsid}. From the comparison of GCR sidereal anisotropy
in different phases from 1999 November to 2008 December, it can be
seen that the CR intensity variation in the local sidereal time
appears fairly stable year by year.

Furthermore, we take a $\chi^2$ test to check the consistency
among different phases. As shown in Equation  (\ref{newchisq}),
the CR intensity variation along the local sidereal time averaged
over all nine phases contains contributions of each single phase.
To avoid the correlation between the two plots which are used in
the test, we compare the result of each single phase with the
average one of all the other eight phases except the single one
instead of the average one of all nine phases. The obtained
$\chi^2/ndf$ value and the
probability of each comparison were labeled in Fig.
\ref{diffphase}. The test also indicates stability for the
sidereal anisotropy with time.

The observation period of Tibet III Array covers more than a half
of the 23rd solar activity cycle from the maximum to the minimum.
So it implies that the sidereal anisotropy of multi-TeV GCRs is
insensitive to the solar activity. It disagrees with the recent result
of Milagro experiment \citep{Milagro2009}, which shows an increase in 
the amplitude of the sidereal anisotropy with time while the phase 
remains stable.

\section{ Conclusions}

In this work, we investigate temporal variations of the large-scale
sidereal anisotropy of GCR intensity using the data of Tibet III Air
Shower Array from 1999 November to 2008 December.
Totally $\sim 4.91\times 10^{10}$ CR events are used.
The data is divided into nine intervals, each in a time span of about
one year. We find that, in the multi-TeV energy range, the sidereal
anisotropy is fairly stable year by year over all nine phases of Tibet
III Array, which covers more than a half of the 23rd solar  cycle
from the maximum to the minimum. It indicates that the
anisotropy in this energy range appears insensitive to solar
activities. This feature can give some constraints on the origin
of the sidereal anisotropy, which has no convincing and widely
accepted explanations so far.

\section{Acknowledgements}

The collaborative experiment of the Tibet Air Shower Arrays has
been performed under the auspices of the Ministry of Science and
Technology of China and the Ministry of Foreign Affairs of Japan.
This work was supported in part by Grants-in-Aid for Scientific
Research on Priority Areas (712) (MEXT), by the Japan Society for
the Promotion of Science (JSPS), by the National Natural Science
Foundation of China, the Chinese Academy of Sciences and the
Ministry of Education of China. C. Fan is partially supported by
the Natural Science Foundation of Shandong Province, China
(No.Q2006A02).


\clearpage

\end{document}